\documentclass[amsmath,amssymb,aps,twocolumn]{revtex4-2}
\usepackage{graphicx}
\usepackage{dcolumn}
\usepackage{bm}
\usepackage{upgreek}
\usepackage[toc,page]{appendix}

\usepackage[normalem]{ulem}

\begin{document}

\preprint{APS/123-QED}

\title{Internal state cooling of an atom with thermal light}

\author{Amanda Younes}
%\email{}
\affiliation{Department of Physics and Astronomy, University of California Los Angeles, Los Angeles, CA, USA}

\author{Randall Putnam}
%\email{}
\affiliation{Department of Physics and Astronomy, University of California Los Angeles, Los Angeles, CA, USA}

\author{Paul Hamilton}
%\email{}
\affiliation{Department of Physics and Astronomy, University of California Los Angeles, Los Angeles, CA, USA}

\author{Wesley C. Campbell}
%\email{}
\affiliation{Department of Physics and Astronomy, University of California Los Angeles, Los Angeles, CA, USA}

\date{\today}

\begin{abstract}

A near-minimal instance of optical cooling is experimentally presented wherein the internal-state entropy of a single atom is reduced more than twofold by illuminating it with broadband, incoherent light.
Since the rate of optical pumping by a thermal state
increases monotonically with its temperature, the cooling power in this scenario increases with higher thermal occupation, an example of a phenomenon known as cooling by heating.
In contrast to optical pumping by coherent, narrow-band laser light, here we perform the same task with fiber-coupled, broadband sunlight, the brightest laboratory-accessible source of continuous blackbody radiation.

\end{abstract}

\maketitle

\section{Introduction}

The study of blackbody radiation (BBR) and its interaction with quantum systems is a subject of research for both fundamental science and for practical applications. BBR plays a critical role in diverse phenomena, including in natural biological processes like photosynthesis, vision, and magnetoreception \cite{Tscherbul2014a,Tscherbul2014,Tscherbul2015, Olsina2014,Brumer2018, Cao2020, Chuang2020,Dodin2021,Chuang2022}. The role of BBR in driving quantum systems is also important in quantum thermodynamics \cite{Scully2011,Dorfman2013, Kosloff2014} and in the fields of spectroscopy and precision measurement, where low-frequency background thermal radiation can shift atomic transition frequencies, limit coherence times, and cause unwanted transitions \cite{Gallagher1979Interactions,Dunbar1991Kinetics,Hechtfischer1998NearThreshold,Beloy2006,Hoekstra2007Optical,Ovsiannikov2011,Safronova2012,Haslinger2018,Lisdat2021,Festa2022BlackbodyRadiation}.

Systems driven by near-resonant BBR can display effects that are often thought of as exclusively quantum mechanical, like steady-state coherence and entanglement, despite interacting only with `classical' thermal light \cite{Koyu2021, Dodin2018, Koyu2022, Dodin2021}. However, the dynamics of quantum systems driven by thermal light have not been thoroughly studied experimentally, and systems of interest are often so large and complicated that they are difficult to study. Therefore, it is useful to consider these interactions in a simple system that is easy to understand and control before studying more complex systems.

One instance of a nontrivial interaction between thermal light interacting and a quantum system is the phenomenon called ``cooling by heating'' \cite{cbh1,cbh2}.  There, the cooling of a system is enhanced by heating an auxiliary quantum mechanical mode that couples the system to a bath, somewhat analogous to a gas-gap heat switch \cite{Pobell}. Cooling by heating schemes have been proposed to cool optomechanical systems \cite{cbh1}, solid-state systems \cite{cbh2}, mechanical resonators \cite{mechres}, and quantum optical systems \cite{qopt}.

Recently, an experimentally accessible scheme for cooling by heating was proposed \cite{firstsunlightpaper} that would use thermal radiation in the visible part of the spectrum to cool a trapped ion to its motional ground state using resolved-sideband cooling \cite{Dehmelt1976Entropy, Wineland1975Proposed, Diedrich1989Laser, Hamann1998Resolved, Teufel2011Sideband}. Here, we demonstrate a variation of that scheme applied only to the internal states of a single trapped ion.  Sunlight, a readily available, high-temperature thermal light source, is coupled into a single mode optical fiber and delivered to a trapped ion to allow a demonstration of a beneficial application of BBR in a controlled quantum experiment. We show internal state cooling by heating of a trapped Ba$^+$ ion via a high-frequency electronic transition driven by sunlight.  We show that the ion's internal state entropy is decreased by more than a factor of two, with cooling power an order of magnitude higher than by spontaneous emission alone.  The production of a low-entropy state in a quantum system by sunlight highlights the potential for quantum effects to manifest in environments common to human experience.

\section{Internal state cooling of a Barium ion}

\begin{figure}[t]
    \centering
    \includegraphics[width = 0.5\textwidth]{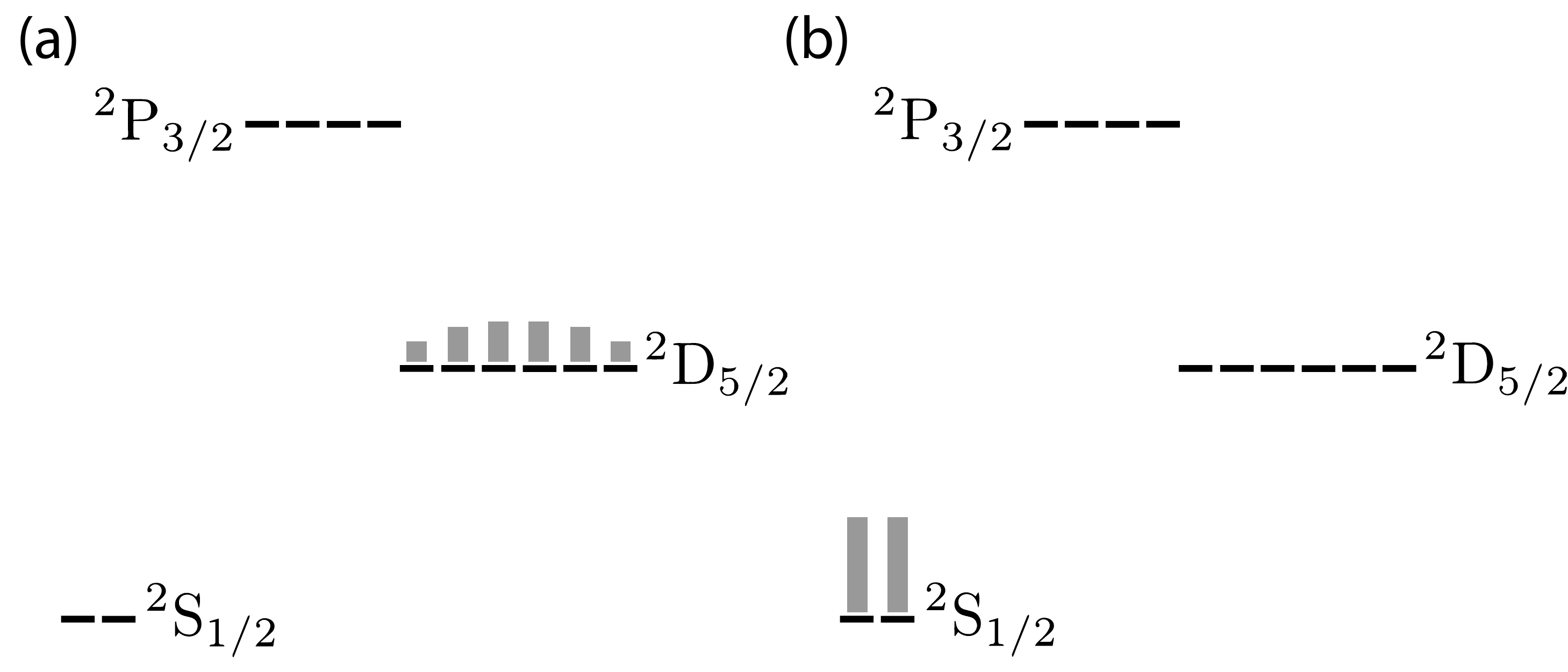}
    \caption{Distribution of population before (a) and after (b) internal state cooling. The ion is prepared in a mixed state with a wide population distribution, corresponding to high von Neumann entropy. After thermal light is applied, the ion is left in the ground state with lower entropy.}
    \label{fig:entropy}
\end{figure}

In contrast to the proposal of Ref.~\cite{firstsunlightpaper}, here we consider cooling of a single valence electron in an atom by considering the magnetic projection levels of an atomic system rather than the atom's center-of-mass motion.  Our experiment focuses on the conversion of a highly-mixed state in the $^2$D$_{5/2}$ manifold of Ba$^+$ to a mixed state in the $^2$S$_{1/2}$ manifold (Fig. \ref{fig:entropy}). Since both mixed states are non-thermal, we monitor the entropy instead of the temperature of the system. The $^2$D$_{5/2}$ manifold contains six magnetic sublevels (projection states), so a maximally-mixed state there has higher entropy, and is thus ‘hotter’, than a maximally mixed $^2$S$_{1/2}$ state. As in sideband cooling, we use thermal light to drive population from all sublevels of the $^2$D$_{5/2}$ manifold to the $^2$P$_{3/2}^o$ manifold, allowing the ion be cooled through spontaneous emission to the ground state. Quantitatively, the von Neumann entropy of maximally mixed state in the metastable manifold has $S=2.58$, while a maximally mixed state in the ground state manifold has $S=1$.

\section{Experimental demonstration}

We choose to demonstrate internal state cooling by heating with sunlight, a readily available source of high-temperature thermal light. Sunlight is coupled with high efficiency into a single mode optical fiber using a home-built tracking device and delivered to the experiment through approximately 200 m of optical fiber \cite{firstsunlightpaper}. This limits the transmitted power spectral density at 614 nm to that of a one-dimensional blackbody at the temperature of the sun \cite{sm_emm, DeVos1988Thermodynamics, Landsberg1989StefanBoltzmann, Alnes2007Blackbody, Fohrmann2015Single}, which is then further suppressed by other factors such as weather conditions, fiber coupling and transmission efficiency, and losses due to focusing optics.

Typical delivery efficiencies at 614 nm are measured to be between 15-40\% of the power spectral density for an ideal blackbody in quasi-one dimension at the temperature of the sun (roughly 5800 K). These factors vary over time and depend on weather conditions, so each measurement is taken at a slightly different power spectral density at the transition frequency.

\begin{figure}[t]
    \centering
    \includegraphics[width = 0.5\textwidth]{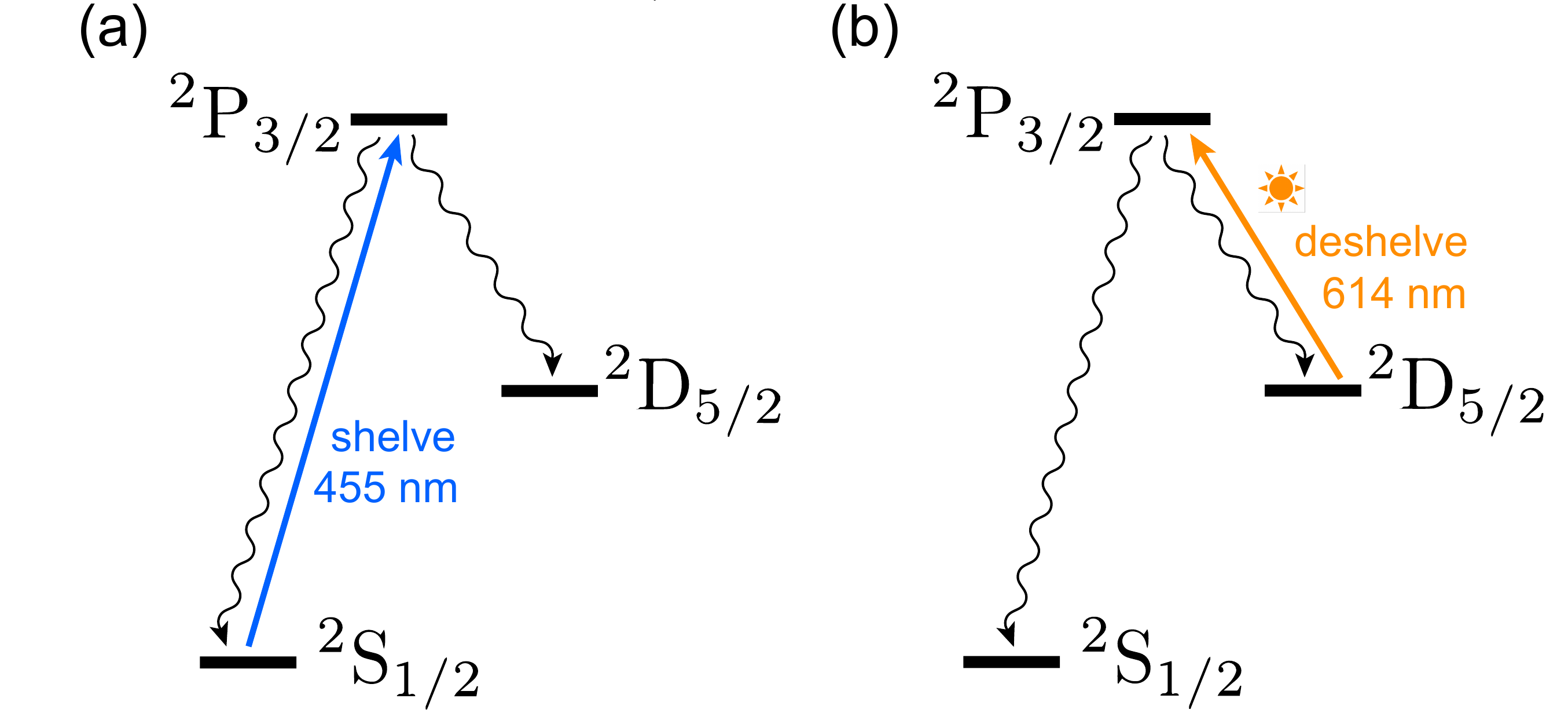}
    \caption{A simplified schematic of the experimental sequence. (a) laser light is used to shelve the ion into the $^2$D$_{5/2}$ manifold then extinguished, followed by (b) the sunlight pumping the ion back to the ground state. We measure the characteristic $1/e$ time for step (b) by illuminating the ion continuously with laser Doppler cooling light while monitoring it for fluorescence. 
    }
   \label{fig:sequence}
\end{figure}

Using this fiber-coupled sunlight, we show internal state cooling by heating of a $^{138}$Ba$^+$ ion in a linear Paul trap. The ion's motion is Doppler cooled and we observe fluorescence on a PMT to monitor the population in the $^2$S$_{1/2}$ state. A small magnetic field of approximately 4.2 G is applied to assist with cooling. Fig. \ref{fig:sequence} shows the experimental sequence. To `shelve' the electron \cite{nagourney_shelved_1986} into the metastable $^2$D$_{5/2}$ state, the ion is illuminated with 455 nm light resonant with the ${}^2\mathrm{S}_{1/2} \leftrightarrow {}^2\mathrm{P}^o_{3/2}$ transition (Fig. [\ref{fig:sequence}(a)]). Once fluorescence disappears, the shelving light is extinguished. We then use unpolarized, unfiltered sunlight to optically pump population back to the $^2$S$_{1/2}$ ground state (Fig. \ref{fig:sequence}(b)). We monitor the fluorescence until we see that population has returned to the ground state, then shelve again and repeat the cycle until a sufficient number of shelve-deshelve cycles have been collected.

This yields a list of dark times which we fit to a decaying exponential function to extract a deshelve time. This method has often been called quantum jump spectroscopy and is commonly used to measure the lifetime of the $^2$D$_{5/2}$ state and other metastable states in trapped ions \cite{madej_quantum_1990, nagourney_shelved_1986}. The sunlight is applied throughout the measurement and has no discernible effect on the ion during the remainder of the sequence.

If the 455 nm laser is polarized so that it includes equal parts $\sigma^+$ and $\sigma^-$ light, as we choose in our experiment, the shelve step prepares a mixed state in the $^2$D$_{5/2}$ with some population in each sublevel. Assuming Doppler cooling produces a mixed state with equal population in the two ground state sublevels, this corresponds to entropy $S=1$. Shelving with this light then produces a state with entropy $S=2.49$. Since the sunlight is unpolarized and propagates approximately perpendicular to the quantization axis, it includes equal parts $\sigma^+$ and $\sigma^-$ light in addition to some $\pi$ light, and can deshelve population in all six sublevels symmetrically, returning the ion to a mixed combination of ground-state sublevels with $S\approx 1$. Deshelving with sunlight in this case decreases the entropy in the internal states of the ion by a factor of almost 2.5.

The above calculation assumes that no quantization field is applied to the ion. However, as stated above, we use an axial field of about 4.2 G, causing a Zeeman splitting between the sublevels. Regardless of the details of the shelve laser intensity and detuning, calculations show that applying sunlight decreases the entropy in the internal states of the ion by approximately the same ratio, and by well over a factor of two.

\subsection{Measured cooling rates}

With no deshelve light at 614 nm and no sunlight on the ion, we expect to measure the natural lifetime of the $^2$D$_{5/2}$ state using the same measurement method. We measure a $1/e$ decay time of $\tau=20.4(5)$ s, compared with a published value of 31.2(9) s \cite{auchter_measurement_2014}, consistent with our estimate ($23-27$ s) based on collisional quenching by background $\mathrm{H}_2$ \cite{madej_quantum_1990, nagourney_shelved_1986}.

\begin{figure}[t]
    \centering
    \includegraphics[width = 0.5\textwidth]{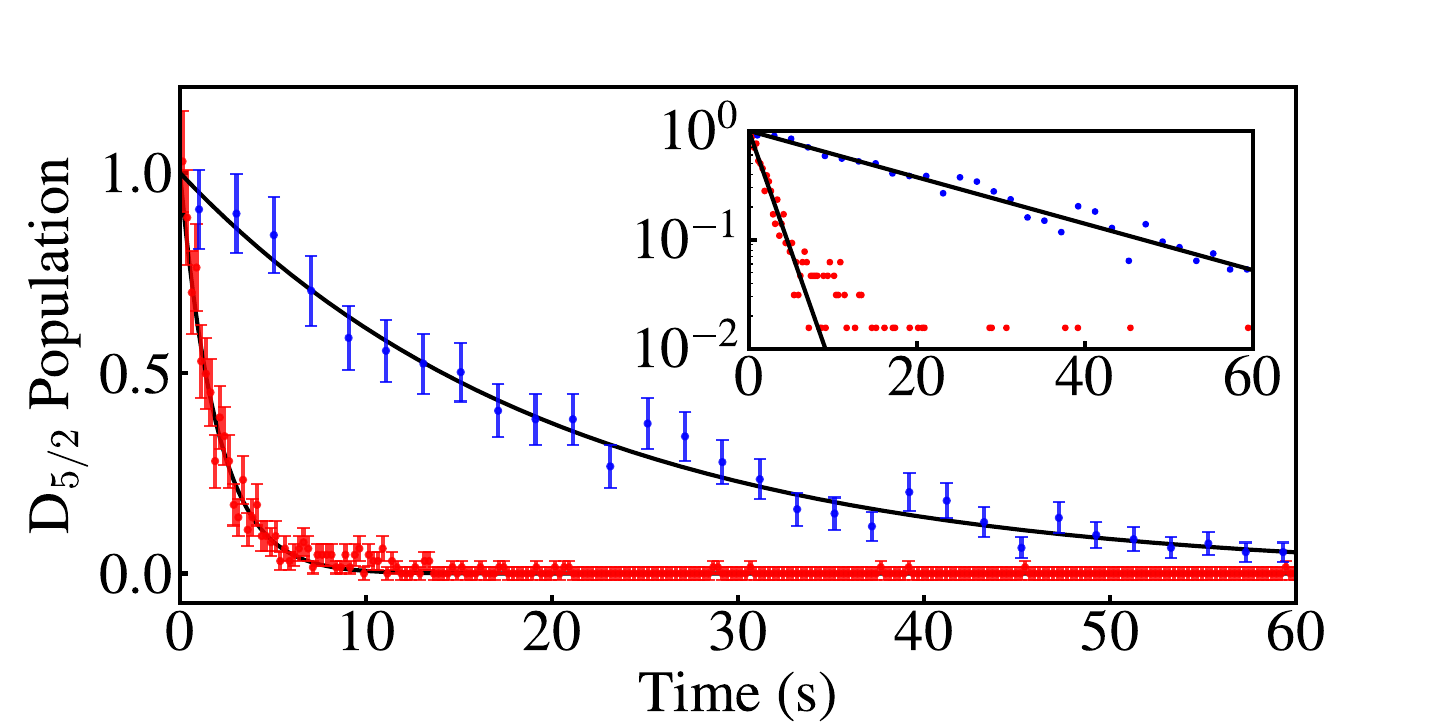}
    \caption{Probability that the ion remains shelved after some time with (red) and without (blue) sunlight. Deshelve times longer than 60 seconds are included in the fit but removed from the plot for readability. Inset is a semilog plot of the same data.}
    \label{fig:compwdark}
\end{figure}

When the 614 nm transition is driven by sunlight, we achieve a deshelve time of $\tau=1.98(5)$ s (Fig. \ref{fig:compwdark}).

We also verify the power spectral density of sunlight delivered to the ion by comparison with excitation of the 614 nm transition with a laser. Despite the difference between excitation with a coherent laser and excitation with incoherent light, the Rabi frequency achieved with an extremely low-intensity 614 nm laser is slow enough compared to the spontaneous emission of the $^2$P$_{3/2}$ state that coherent effects disappear almost immediately, and the laser excitation behaves similarly to incoherent driving of the transition. Therefore, the spectral density of the sunlight at 614 nm can be directly compared to the spectral density of the laser by comparing their transition rates.

The deshelve laser is passed through the same optical path and focusing system as the sunlight and detuned 50 GHz from the deshelve transition, to assist in slowing down the deshelve rate and to compensate for a broad laser linewidth. We then measure a deshelve time using the same method as previously, for multiple laser powers that yield a deshelve time on the order of a few seconds. These times are fit to a theoretical result for the deshelve operation with a far detuned laser, and an effective area is determined which incorporates the beam size and any factors due to polarization of the shelve and deshelve light. We then convert to a resonant power spectral density at 614 nm that would yield the same deshelve time as the measured value with sunlight. This yields a power spectral density of $7.8$ nW/THz which is within a factor of 3 of the inferred value given our measured deshelve time with sunlight ($2.9$ nW/THz), and is consistent with measured delivery efficiencies.

\begin{acknowledgments}
The authors acknowledge Kristian Barajas for discussions and calculations. This work was supported by the NSF under PHY-2207985 and OMA-2016245.
\end{acknowledgments}

\bibliography{refs}

\end{document}